\documentclass[11pt]{article}

\usepackage{amsmath}
\usepackage{amssymb}
\usepackage{graphicx}
\usepackage{subfigure}

\def\middlespace {\smallskipamount=5.625pt plus1.5pt minus1.5pt
                  \medskipamount=11.25pt plus3pt minus3pt
                  \bigskipamount=22.5pt plus6pt minus6pt
                  \normalbaselineskip=22.5pt plus0pt minus0pt
                  \normallineskip=1pt
                  \normallineskiplimit=0pt
                  \jot=5.625pt
                  {\def\smallskip {\vskip\smallskipamount}}
                  {\def\medskip   {\vskip\medskipamount}}
                  {\def\bigskip   {\vskip\bigskipamount}}
                  {\setbox\strutbox=\hbox{\vrule
                    height15.75pt depth6.75pt width 0pt}}
                  \parskip 11.25pt
                  \normalbaselines}

\begin{document}

\ \vskip 1.0 in

\begin{center}
 { \Large {\bf Modified Gravity as a Common Cause for  }}

\smallskip

{\Large {\bf  Cosmic Acceleration and Flat Galaxy Rotation Curves}}

\vskip 0.2 in

\smallskip

\bigskip

\bigskip

\bigskip

{{\large
{\bf Priti Mishra and Tejinder P. Singh 
} 
}}

\medskip

{\it Tata Institute of Fundamental Research,}\\
{\it Homi Bhabha Road, Mumbai 400 005, India}\\
\smallskip
{\tt e-mail addresses: priti@tifr.res.in, tpsingh@tifr.res.in}\\
\bigskip
%{\it 22nd March, 2010}
\vskip 0.5cm
\end{center}

\vskip 1.0 in

\begin{abstract}

\noindent Flat galaxy rotation curves and the accelerating Universe both imply the existence of a critical acceleration, which is of the same order of magnitude in both the cases, in spite of the galactic and cosmic length scales being vastly different. Yet, it is customary to explain galactic acceleration by invoking gravitationally bound dark matter, and cosmic acceleration by invoking a `repulsive` dark energy. Instead, might it not be the case that the flatness of rotation curves and the acceleration of the Universe have a common cause? In this essay we propose a modified theory of gravity. By applying the theory on galactic scales we demonstrate flat rotation curves without dark matter, and by applying it on cosmological scales we demonstrate cosmic acceleration without dark energy.

\bigskip

\bigskip

\centerline{March 22, 2012}

\noindent 

\noindent 

\vskip 1.0 in

% submited to GRF 22 March, 2012; arXiv version
\centerline{\it Essay written for the Gravity Research Foundation 2012 Awards for Essays on Gravitation}
\centerline{\it This essay received an Honorable Mention}
\smallskip
\centerline{Published Online in Int. J. Mod. Phys. 21 (2012) 1242002 (5 pages)  DOI: 10.1142/S0218271812420023}    
\end{abstract}

\newpage

\middlespace

\noindent Dark matter has been postulated to explain the flat non-Keplerian behaviour of rotation curves of galaxies, and the high dispersion velocities of galaxy clusters. Dark energy/cosmological constant has been postulated as an explanation for the observed late-time cosmic acceleration. Together, cold dark matter and a cosmological constant provide the standard $\Lambda CDM$ model of structure formation and large-scale-structure which best fits the current cosmological data. Nonetheless, there are chinks in the armour! Apart from the well-known facts that dark matter has not yet been detected in the laboratory, and that the inferred value of the cosmolgical constant is too low compared to the theoretically preferred value by some $10^{120}$ orders of magnitude, there is another remarkable observation which the $\Lambda CDM$ model does not seem to account for. And that is the existence of a universal acceleration 
$cH_0 \sim  10^{-10}$ cm sec$^{-2}$, where $H_0$ is the present value of the Hubble parameter, which prevails in the outer regions of galaxies, in galaxy clusters, and at the edge of the observed Universe:
\begin{equation}
\frac{GM_{galaxy}}{R_{galaxy}^2}\quad \sim  \quad \frac{GM_{cluster}}{R_{cluster}^2} \quad\sim\quad \frac{GM_{Universe}}{R_{Universe}^2} \quad\sim \quad cH_0
\label{uniacc}
\end{equation}

In this essay, we seek an understanding of the cause for this universal acceleration, and for flat galaxy rotation curves and cosmic acceleration, by postulating a fourth order modified gravity as a common  explanation, as an alternative to dark matter and dark energy. We propose, for reasons to be discussed below, that the current structure and evolution of the Universe is described by the following effective field equations
\begin{equation}
R^{\mu\nu} - \frac{1}{2} g^{\mu\nu} R = \frac{8\pi G}{c^4} T^{\mu\nu} + 
L^{2} R^{\mu\alpha\nu\beta}_{\ \ \ \ \ ;\alpha\beta}
\label{modee}
\end{equation}

The length parameter $L$ is scale-dependent and defined as being proportional to the square-root of the mass $M$ of the system under study, in such a way that $GM/L^2$ is a universal constant equal to $cH_0$. This holographic assumption [mass proportional to area rather than volume] seems to be universally valid for large astrophysical systems (see e.g. Fig. 7 in ~\cite{Moffat:2009}) and entirely consistent with the observation in Eqn. (\ref{uniacc}). 

We will solve these modified gravity equations for two different spacetime metrics:

(i) The Newtonian weak-field non-relativistic approximation for the scalar potential $\phi$: 
\begin{equation}
ds^2 =  \left( 1 + \frac{2\phi}{c^2}\right) c^{2} dt^2 - dx^2 - dy^2 - dz^2\ ;
\label{weakfd}
\end{equation}
and 

(ii) The spatially flat Robertson-Walker cosmological metric
\begin{equation}
ds^2 =  c^{2} dt^{2}   - a^{2}(t) [dx^2 + dy^2 + dz^2]\; .
\label{rw}
\end{equation} 
In the first case we will show how the resulting fourth order biharmonic modification of the Poisson equation explains flat galaxy rotation curves without dark matter, and in the second case we will show how a fourth order modification of the Friedmann equations explains late-time cosmic acceleration without dark energy. 

It can be shown that for the weak-field metric (\ref{weakfd}) the modified equations (\ref{modee}) reduce to the following fourth order biharmonic correction to the Poisson equation ~\cite{Zala, Priti}
\begin{equation}
\nabla^2\phi - L^2 \nabla^4\phi =4\pi G\mu(r)
\label{biharmonic}
\end{equation}
where $\mu(r)$ is the matter density distribution of interest. From observations of galaxies this is known to be 
\begin{equation}
\mu(r)=\frac{3}{4\pi r^{3}} \beta  M(r) \bigg[\frac{r_c}{r+r_c}\bigg]
\end{equation}
where
\begin{equation}
M(r)=4\pi \int_0^rdr'r'^2\mu(r') = M \left( \frac{r}{r+r_{c}}\right)^{3\beta}
\end{equation}
and
\begin{equation}
\beta = \left\{
\begin{array}{ll} 1 & \mbox{for HSB galaxies,} \\
2 & \mbox{for LSB \& Dwarf galaxies.}
\end{array} \right.
\end{equation}

For this mass distribution, the biharmonic equation (\ref{biharmonic}) can be solved to obtain  the following remarkable Yukawa form for the radial component of the gravitational acceleration $a=-\nabla\phi$ inside the galaxy
\begin{equation}
a(r)=-\frac{GM(r)}{r^2}\bigg\{1+\sqrt{\frac{M_0}{M}}\bigg[1-\exp(-r/L)\bigg(1+\frac{r}{L}\bigg)\bigg] \bigg\}
\label{yukawa}
\end{equation}
where $M_0$ is a constant of integration and $M$ is the total mass of the galaxy ~\cite{Priti}. The rotation curve $v(r)$  of the galaxy is obtained by setting $a(r) = v^{2}(r)/r$. If we choose $L$ according to the prescribed relation $GM/L^2 = cH_0$ [this gives $L\sim 10$ kpc] and if we choose $M_0 = 10^{12}M_{\odot}$ it is possible to fit the flat rotation curves of a large class of galaxies, without invoking dark matter ~\cite{Priti, Brownstein:2006}. It is to be noted that the values of $L$ and $M_0$ chosen here are consistent with all known laboratory and astronomical tests of departures from the inverse square 
law ~\cite{adel}. For $r\ll L$, the inverse square law holds. Also, for $r\gg L$ the inverse square law holds but with an effectively larger value of the gravitational constant. Interesting new physics arises around 
$r\gtrsim L$.

Greater insight into the modified acceleration obtained by us can be had by expanding the acceleration (\ref{yukawa}) around $r=L$ and writing it as a sum of two terms : a part that falls as $1/r^2$ and is independent of $L$, and a part that falls as $1/r$ and depends on $L$. This gives

\begin{equation}
a(r)\approx - \frac{GM(r)}{r^2}\bigg[1+\sqrt{\frac{M_0}{M}}\bigg\{1-\frac{3}{e}\bigg\}\bigg]- \frac{GM(r)}{r}\bigg[
\sqrt{\frac{M_0}{M}}\frac{1}{L e}\bigg].
\label{accelrationintwoparts}
\end{equation}

\begin{figure}[!ht]
\begin{center}
\includegraphics[width=\textwidth]{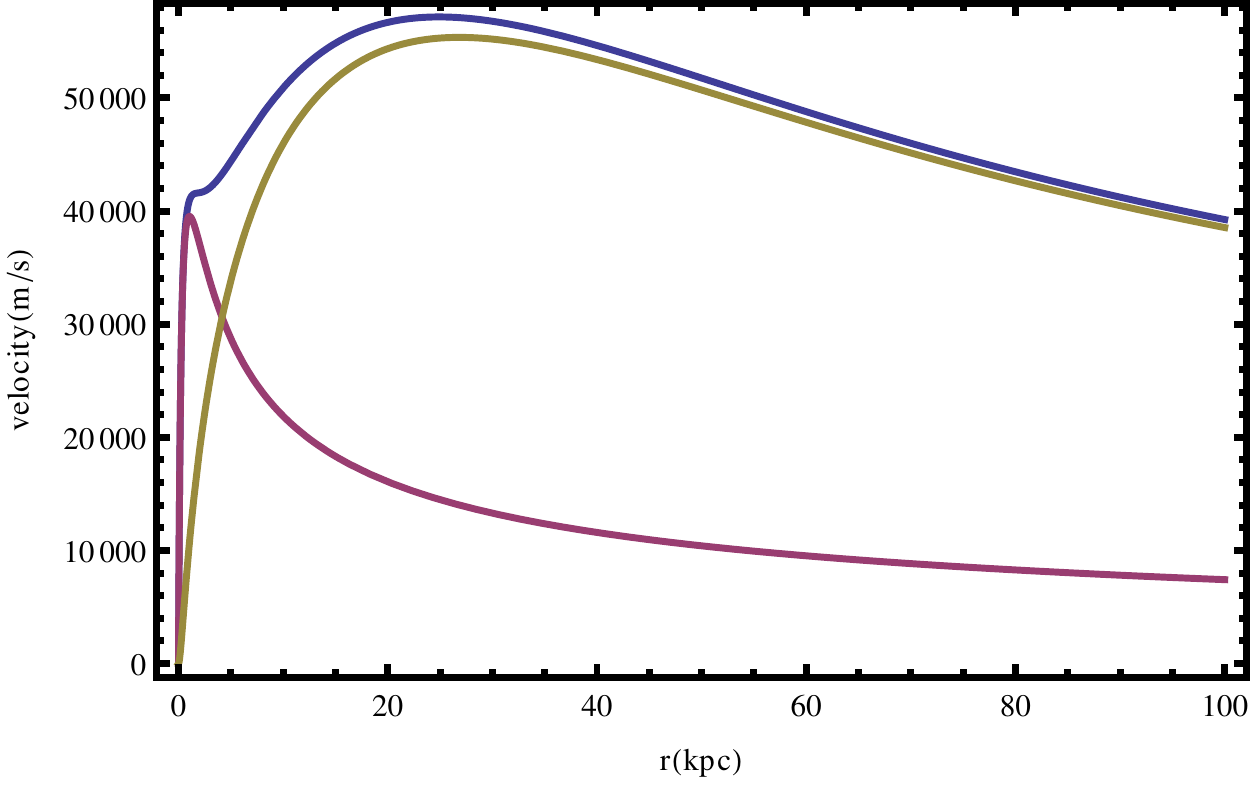}
\caption{Red curve: velocity due to first term, Green curve: velocity due to second term, 
Blue curve: total velocity.}
\label{total velocity}
\end{center}
\end{figure}

The second term dominates for $r\gtrsim L$. Fig. 1 above plots the rotation velocity curve, and it is clear that the Keplerian fall-off due to conventional Newtonian acceleration [first term] is modified to a flat rotation curve [second term] for $r\gtrsim L$, because of the proposed generalisation of Einstein gravity. 

Next, we turn to cosmology and write the modified Friedmann equations resulting from Eqn. (\ref{modee}), for the Robertson-Walker metric. Since one is no longer in the weak-field limit, the equations are complicated, but it is adequate for our purpose  to write them in the form
\begin{equation}
\frac{\dot{a}^2}{a^2} - L_{U}^2 {\cal F}_1(a,\dot a, \ddot{a}, \dddot{a}) = \frac{8\pi G}{3}\rho
\label{density}
\end{equation}
\begin{equation}
\frac{2\ddot a}{a} + \frac{{\dot a}^2}{{\dot a}^2} - L_{U}^2 {\cal F}_{2}(a,\dot{a},\ddot{a},\dddot{a}, \ddddot{a})=
8\pi G p
\label{accn}
\end{equation}
where ${\cal F}_1$ and ${\cal F}_2$ are known functions of their arguments which we do not write explicitly. In accordance with the scaling principle, since we are now considering the entire observable Universe, we have $L_U = cH_{0}^{-1}$. It turns out that these equations possess a very interesting analogy with the weak-field case considered above. There is a space $\leftrightarrow$ time symmetry: the behaviour seen in the previous case around $r\gtrsim L$ is repeated here for $t\gtrsim H_{0}^{-1}$. 

For times $t \ll L_U$ the modifying gravity terms can be neglected and these equations reduce to the standard Friedmann equations which give a decelerating solution. This is analogous to the behaviour seen in the solution (\ref{yukawa}) for the gravitational acceleration: just as the inverse-square law holds for $r \ll L$ in Eqn. (\ref{yukawa}), for $t\ll H_{0}^{-1}$ one recovers the standard Friedmann evolution. Things start to get interesting for $t\gtrsim L_U = H_{0}^{-1}$. For the sake of simplification one can divide the evolution into two phases: (i) evolution for $t\ll H_{0}^{-1}$, and (ii) evolution for $t \gtrsim H_{0}^{-1}$. In the second case, one can show from Eqn. (\ref{accn}) for the dust case $p=0$, assuming a power-law solution, that the evolution is dominated by the modifying terms, and the scale-factor is given by
\begin{equation}
a(t)  = \left ( \frac{t-L_U}{T-L_U}\right)^{3/2} 
\label{phase2}
\end{equation}
where $T$ is the current age of the universe. One has an accelerating solution with $\ddot{a}>0$. This is in complete analogy with the galactic case: the modifying term increases the acceleration, and in the present cosmological case the rise is enough to change the sign from deceleration to postive acceleration. Furthermore, the acceleration decreases with increasing time, and it can be shown that for $t\gg H_{0}^{-1}$ the decelerating phase reappears. It is easily shown that $T=5H_{0}^{-1}/2$, and as can be expected the acceleration is of the order of $cH_0$. [The vanishing of the scale factor at $t=L_U$ is only an approximation, and simply reflects the smallness of its value at $t=L_U$ compared to the present value $a=1$; the true singularity of course occurs at the beginning of the previous decelerating phase].

Using the solution (\ref{phase2}) for the scale factor in the first Friedmann equation (\ref{density}) it can be shown that the dust density evolves as
\begin{equation}
\rho(t) \propto \frac{1}{t^2} \propto \frac{1}{a^{4/3}}.
\end{equation}
The density falls with the scale factor slowly compared to the standard case [where $\rho\propto 1/a^3$], because the modifying term in the Einstein equation implies that
\begin{equation}
 \left( \frac{8\pi G}{c^4} T^{\mu\nu} + 
L^{2} R^{\mu\alpha\nu\beta}_{\ \ \ \ \ ;\alpha\beta} \right)_{;\nu} = 0.
\end{equation}
The onset of fourth order gravity at the present cosmological epoch, which we have proposed  as a phenomenological representation of an underlying physics, implies an exchange of energy-momentum between gravity and matter fields at galactic and cosmological scales. This is one possible unified way of understanding the otherwise unusual features of flat rotation curves and cosmic acceleration.

The underlying physics may possibly have to do with the effect that averaging over small-scale 
inhomogeneities has on dynamics on larger scales, a phenomenon which is perhaps not fully understood at present. In fact, such considerations led to the suggestion from Szekeres ~\cite{szekeres} of essentially the same equation as Eqn. (\ref{modee}) proposed here.  

Understanding structure formation in this picture is a challenge.  But we have a unified description of critical acceleration on different scales, which comes, not from invoking different matter/energy components, but a phenomenological modification of gravity whose root cause may be the same at different scales, and which calls for further study.

\end{document}